 \pdfoutput=1
\documentclass[letterpaper, 10 pt, conference]{ieeeconf}  

\IEEEoverridecommandlockouts                              

\usepackage{amsmath} 
\usepackage{tikz}
\usepackage{physics}
\usepackage{subcaption}
\usetikzlibrary{patterns}
\usetikzlibrary{decorations.pathreplacing,angles,quotes}

\usepackage[version=4]{mhchem}
\usepackage{nomencl}
\usepackage{etoolbox}
\renewcommand\nomgroup[1]{%
  \item[\bfseries
  \ifstrequal{#1}{N}{Notation}{%
  }%
  \ifstrequal{#1}{S}{Sub- \&Superscripts}{%
  }%
  \ifstrequal{#1}{G}{Geometry}{%
  }%
  \ifstrequal{#1}{M}{Math}{%
  }%
  \ifstrequal{#1}{P}{Physics}{%
  }%
]}

\makenomenclature

\title{\LARGE \bf
A Dynamic Cooler Model for Cement Clinker Production
}

\author{Jan Lorenz Svensen$^{1,2}$, Wilson Ricardo Leal da Silva$^{2}$, Javier Pigazo Merino$^{2}$,\\ Dinesh Sampath$^{2}$  and John Bagterp J\o rgensen$^{1}$
\thanks{*This work was supported by Innovation Fund Denmark, Ref. 2053-00012B}
\thanks{$^{1}$ DTU Compute, Department of Applied Mathematics and Computer Science, Technical University of Denmark, 2800 Lyngby, Denmark
        {\tt\small jlsv@dtu.dk, jbjo@dtu.dk}}%
\thanks{$^{2}$ FLSmdth A/S, 2500, Valby, Denmark
        {\tt\small wld@flsmidth.com}}%
}

\begin{document}

\maketitle
\thispagestyle{empty}
\pagestyle{empty}

\begin{abstract}
We present a 2D model for a grate belt cooler in the pyro-section of a cement plant.
The model is formulated as an index-1 differential-algebraic equation (DAE) model based on first engineering principles. The model systematically integrates thermo-physical aspects, transport phenomena, reaction kinetics, mass and energy balances, and algebraic volume and energy relations.
The model is used for dynamic simulation of the cooler and the paper provides dynamic and steady-state simulation results matching the expected behavior. 
The cooler model is one part of a full pyro-section model for dynamical simulations.
The model can serve as a basis for the design of optimization and control systems towards improving energy efficiency and \ce{CO2} emission.

\end{abstract}
\section{Introduction}
The manufacturing of cement corresponds to 8\% of the global \ce{CO2} emissions \cite{CO2Techreport}. A main source of these emissions comes from the production of cement clinker.
On the transition towards zero \ce{CO2} emissions, optimization, control, and digital solutions are important approaches just as process modification for carbon capture and alternative materials are. 
For the development of digital, control, and optimization approaches for the pyro-process, and the cement plants in general, dynamic simulations are a requirement.

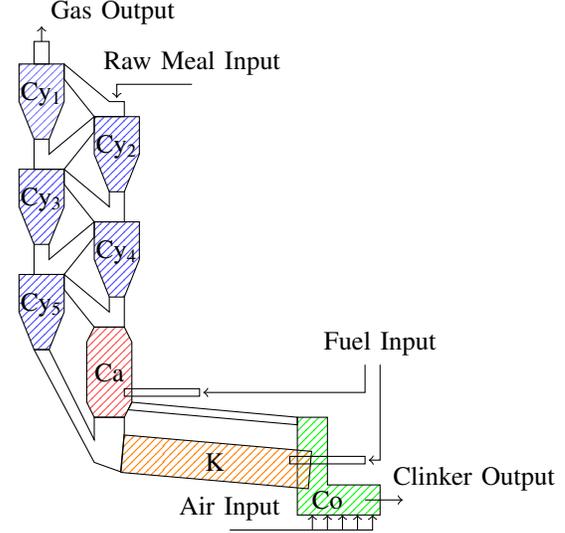
\begin{figure}[tb]
    \centering
        \begin{tikzpicture}    
            \draw[pattern=north east lines, pattern color=blue!50] (0.2,5.5) -- (0.2,6) -- (0.8,6) -- (0.8,5.5) -- (0.6,5) -- (0.4,5)-- cycle;
            \draw[pattern=north east lines, pattern color=blue!70] (1.2,4.8) -- (1.2,5.3) -- (1.8,5.3) -- (1.8,4.8) -- (1.6,4.3) -- (1.4,4.3)-- cycle;
            \draw[pattern=north east lines, pattern color=blue!70] (0.2,4.1) -- (0.2,4.6) -- (0.8,4.6) -- (0.8,4.1) -- (0.6,3.6) -- (0.4,3.6)-- cycle;
            \draw[pattern=north east lines, pattern color=blue!70] (1.2,3.4) -- (1.2,3.9) -- (1.8,3.9) -- (1.8,3.4) -- (1.6,2.9) -- (1.4,2.9)-- cycle;
            \draw[pattern=north east lines, pattern color=blue!70] (0.2,2.7) -- (0.2,3.2) -- (0.8,3.2) -- (0.8,2.7) -- (0.6,2.2) -- (0.4,2.2)-- cycle;

            \draw  (1.4,5.5) -- (0.8,6) -- (0.8,5.8) -- (1.2,5.3)-- (1.6,5.3) -- (1.6,5.5) -- cycle;
            \draw (0.6,5) -- (0.6,4.8) -- (1.2,5.3) -- (1.2,5.1) -- (0.8,4.6)-- (0.4,4.6) -- (0.4,5);
            \draw (1.4,4.3) -- (1.4,4.1) -- (0.8,4.6) -- (0.8,4.4) -- (1.2,3.9)-- (1.6,3.9) -- (1.6,4.3);
            \draw (0.6,3.6) -- (0.6,3.4) -- (1.2,3.9) -- (1.2,3.7) -- (0.8,3.2)-- (0.4,3.2) -- (0.4,3.6);
            \draw (1.4,2.9) -- (1.4,2.7) -- (0.8,3.2) -- (0.8,3.0) -- (1.2,2.5)-- (1.6,2.5) -- (1.6,2.9);
            \draw (0.5,2.8) node{\ce{Cy5}};
            \draw (1.5,3.5) node{\ce{Cy4}};
            \draw (0.5,4.2) node{\ce{Cy3}};
            \draw (1.5,4.9) node{\ce{Cy2}};
            \draw (0.5,5.6) node{\ce{Cy1}};
            
            \draw[pattern=north east lines, pattern color=red!70] (1.1,1.5) -- (1.1,2.3) -- (1.2,2.5) -- (1.6,2.5) -- (1.7,2.3) -- (1.7,1.5) -- (1.6,1.3) -- (1.2,1.3)-- cycle;
            \draw (1.4,1.9) node{Ca};
            
            \draw[pattern=north east lines, pattern color=green] (3.9,1+0.3) -- (3.9,1-1) -- (5,1-1) -- (5,1-0.6) -- (4.3,1-0.6) -- (4.3,1+0.3)-- cycle;
            \draw[rotate around={-5:(0,0)},pattern=north east lines, pattern color=orange] (1.5,1-0.3) -- (4,1-0.3) -- (4,1+0.2) -- (1.5,1+0.2) -- cycle;
            \draw (2.8,0.7) node{K};
            
            \draw (3.9,1+0.3) -- (3.9,1+0.2) -- (1.65,1.4) -- (1.7,1.5) -- cycle;
            \draw (1.6,1.3) -- (1.6,1.1) -- (1.55,0.575) -- (1.2,0.7) -- (0.4,2.2) -- (0.6,2.2) -- (1.2, 1) -- (1.2,1.3) -- cycle;
            \draw (4.3,0.2) node{Co};

            \draw  (3.8,0.68) rectangle (4.8,0.78);
            \draw  (1.6,1.58) rectangle (2.6,1.68);
            \draw  (0.4,6) rectangle (0.6,6.3);

            \draw[->] (4.8,0.2) -- node[anchor=south west]{Clinker Output} (5.3,0.2);
            \draw[->] (5.0,2) node[anchor=south]{Fuel Input} -- (5.0,0.73)  -- (4.85,0.73);
            \draw[->] (4.8,2) -- (4.8,1.63)  -- (2.65,1.63);
            \draw[->] (0.5,6.3) -- node[anchor=south west]{Gas Output} (0.5,6.5);
            \draw[->] (2.5,5.73) node[anchor=south]{Raw Meal Input} -- (1.5,5.73)  -- (1.5,5.55);
            \draw[->] (3.0,-0.2) node[anchor=south]{Air Input} -- (4.1,-0.2)  -- (4.1,0.0);
            \draw[->] (4.1,-0.2) -- (4.3,-0.2)  -- (4.3,0.0);
            \draw[->] (4.3,-0.2) -- (4.5,-0.2)  -- (4.5,0.0);
            \draw[->] (4.5,-0.2) -- (4.7,-0.2)  -- (4.7,0.0);
            \draw[->] (4.7,-0.2) -- (4.9,-0.2)  -- (4.9,0.0);
    \end{tikzpicture}    
    \caption{
    The pyro-section for clinker production in a cement plant consists of preheating tower of cyclones ({\color{blue}Cy}), a calciner ({\color{red}Ca}), a rotary kiln ({\color{orange}K}), and a cooler ({\color{green}Co}).
    }\label{fig:production}
\end{figure}

Fig. \ref{fig:production} illustrates the pyro-process of cement clinker production. The process consists of a pre-heating tower of cyclones, a calciner, a rotary kiln, and a cooler.
In this paper, we provide a mathematical model for dynamic simulations
of the cooler. We specifically consider a grate belt cooler.
The cooler model is useful for the design of control and optimization systems for both heat recirculation and clinker quality, critical for the operation of cement plants.

In the existing literature on models, Metzger modeled the cooler as a dynamic 1D model of energy balances without a radiation term \cite{METZGER1983491}. Cui et al suggested a steady-state 2D model for the energy balances with porosity, though without radiation or mass balances \cite{CUI20171297}. Mujumdar et al suggested a steady-state 2D model for the energy balance with applied porosity and a 1D mass balance model for the overall mass flow \cite{MUJUMDAR2007}. 

In comparison, we provide a mathematical 2D model for dynamic simulations of the cooler, including both mass and energy balances for the material and gas content. The model describes the material compositions, flows, and heat exchanges based on rigorous thermo-physical properties and kinetic expressions.
The model is formulated as a system of index-1 differential algebraic equations (DAEs), using a novel systematic modeling methodology that integrates thermo-physical properties, transport phenomena, and stoichiometry and kinetics with mass and energy balances. 
The modeling approach allows for simple formulations of models without the need of supplementary assumptions, e.g. assuming properties like heat capacity being constant.
Models of the complementary parts of the pyro-process can be found in \cite{Svensen2024Kiln} and \cite{Svensen2024Calciner} for rotary kiln and calciner, respectively. The literature provides references on industrial cooler operation and design practices \cite{bhatty2010innovations}\cite{chatterjee2018cement}\cite{bye1999portland}.


This paper is organized as follows. Section \ref{sec:Cooler} presents the cooler, while Section \ref{sec:CoolerModel} describes the mathematical model for the cooler. Section \ref{sec:SimulationResults} presents simulation results and Section \ref{sec:Conclusion} provides the conclusions of the paper.
\section{The Cooler}\label{sec:Cooler}
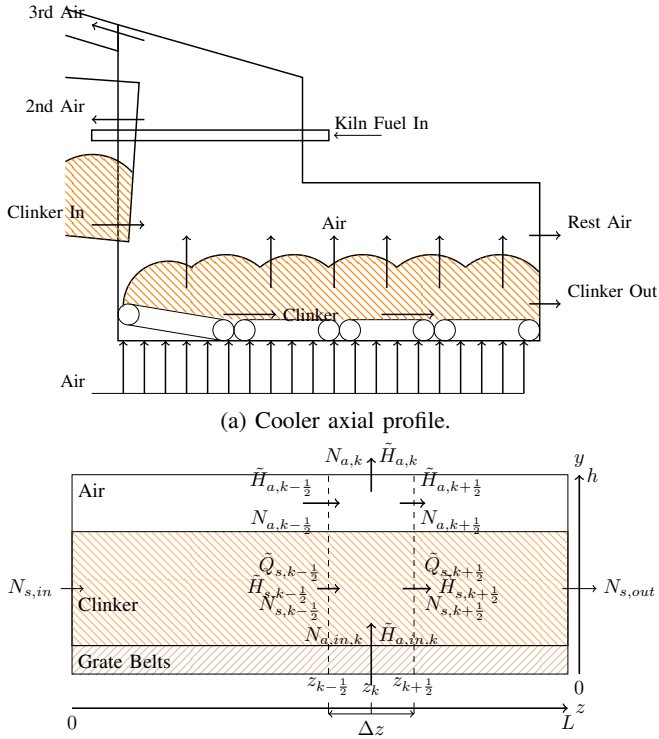
\begin{figure}
    \centering
\begin{subfigure}[b]{0.5\textwidth}    
    \centering
    \resizebox{1\textwidth}{!}{%
    \begin{tikzpicture}   
        \draw[thick] (0,0)  -- (0,3) -- (0,6) -- (3.5, 5) -- (3.5,3) -- (8,3) -- (8,0)-- cycle;
        \draw[thick] (-0.5,3.8) rectangle (4,4);
        \draw[thick] (-1,6.3) -- (0,6) -- (0,5.5) -- (-1,5.8);
        \draw[thick] (-1,5.0) -- (0.4,4.9) -- (0.2,1.88) -- (-1,2.0);
        
        \draw[-] (0.2,0.5) circle (0.2 cm);        
        \draw[-] (2.0,0.2) circle (0.2 cm);
            \draw[-] (0.2,0.7) -- (2.0,0.4);
            \draw[-] (0.2,0.3) -- (2.0,0.0);
        \draw[-] (2.4,0.2) circle (0.2 cm);
        \draw[-] (4.0,0.2) circle (0.2 cm);
            \draw[-] (2.4,0.4) -- (4.0,0.4);
        \draw[-] (4.4,0.2) circle (0.2 cm);
        \draw[-] (5.8,0.2) circle (0.2 cm);
            \draw[-] (4.4,0.4) -- (5.8,0.4);
        \draw[-] (6.2,0.2) circle (0.2 cm);
        \draw[-] (7.8,0.2) circle (0.2 cm);
            \draw[-] (6.2,0.4) -- (7.8,0.4);

        \draw[thick,-] (0.1,0.65) arc (180:58:0.86cm);
        \draw[thick,-] (1.4,1.4) arc (130:49:1cm);
        \draw[thick,-] (2.7,1.4) arc (130:49:1cm);
        \draw[thick,-] (4.0,1.4) arc (130:49:1cm);
        \draw[thick,-] (5.3,1.4) arc (130:49:1cm);
        \draw[thick,-] (6.6,1.4) arc (130:42:1cm);
        
        \draw[thick,-] (-1.0,3.4) arc (120:42:1.02cm);
        \fill[pattern=north west lines, pattern color=orange] (-1.0,3.4) arc (120:42:1.02cm) -- (0.2,1.88) -- (-1,2.0) -- cycle;
        \fill[pattern=north west lines, pattern color=orange] (0.1,0.65) arc (180:58:0.86cm) -- (1.4,1.4) arc (130:49:1cm)-- (2.7,1.4) arc (130:49:1cm)-- (4.0,1.4) arc (130:49:1cm)-- (5.3,1.4) arc (130:49:1cm)-- (6.6,1.4) arc (130:42:1cm) -- (8,0.4) -- (2.2,0.4) -- (0.2,0.7)-- cycle;

        \draw[-] (-0.5,-1)node[anchor=south east]{Air} -- (7.7,-1);
          \draw[thick,->] (0.1,-1) -- (0.1,0);  
          \draw[thick,->] (0.5,-1) -- (0.5,0);  
          \draw[thick,->] (0.9,-1) -- (0.9,0);  
          \draw[thick,->] (1.3,-1) -- (1.3,0);  
          \draw[thick,->] (1.7,-1) -- (1.7,0);  
          \draw[thick,->] (2.1,-1) -- (2.1,0);  
          \draw[thick,->] (2.5,-1) -- (2.5,0);  
          \draw[thick,->] (2.9,-1) -- (2.9,0);  
          \draw[thick,->] (3.3,-1) -- (3.3,0);
          \draw[thick,->] (3.7,-1) -- (3.7,0);  
          \draw[thick,->] (4.1,-1) -- (4.1,0);  
          \draw[thick,->] (4.5,-1) -- (4.5,0);  
          \draw[thick,->] (4.9,-1) -- (4.9,0);  
          \draw[thick,->] (5.3,-1) -- (5.3,0);  
          \draw[thick,->] (5.7,-1) -- (5.7,0);  
          \draw[thick,->] (6.1,-1) -- (6.1,0);  
          \draw[thick,->] (6.5,-1) -- (6.5,0);  
          \draw[thick,->] (6.9,-1) -- (6.9,0);  
          \draw[thick,->] (7.3,-1) -- (7.3,0);
          \draw[thick,->] (7.7,-1) -- (7.7,0); 

          \draw[thick,->] (1.3,1) -- (1.3,2);
          \draw[thick,->] (2.9,1) -- (2.9,2); 
          \draw[thick,->] (4.1,1) -- (4.1,2)node[anchor=south]{Air};  
          \draw[thick,->] (5.1,1) -- (5.1,2);
          \draw[thick,->] (6.1,1) -- (6.1,2);  
          \draw[thick,->] (7.3,1) -- (7.3,2);

          \draw[thick,->] (2,0.5) -- (3,0.5)node[anchor=west]{Clinker};
          \draw[thick,->] (5,0.5) -- (6,0.5);

          \draw[thick,->] (7.8,0.7)--(8.4,0.7) node[anchor=south west]{Clinker Out};

          \draw[thick,->] (-0.5,2.2)node[anchor=south east]{Clinker In}--(0.5,2.2) ;
          \draw[thick,<-] (-0.5,4.2)node[anchor=south east]{2nd Air}--(0.5,4.2) ;
          \draw[thick,<-] (-0.5,6.0)node[anchor=south east]{3rd Air}--(0.5,5.7) ;
          \draw[thick,<-] (8.4,2.0)node[anchor=south west]{Rest Air}--(7.8,2.0) ;

          \draw[<-] (4.1,3.9) -- (5,3.9) node[anchor=south]{Kiln Fuel In};
    \end{tikzpicture}
     }
    \caption{Cooler axial profile.}
    \label{fig:cooler}
\end{subfigure}

\begin{subfigure}[b]{0.5\textwidth}    
    \centering
     \resizebox{1\textwidth}{!}{%
     \begin{tikzpicture}   
        \draw[thick,->] (0,-0.6)node[anchor=north]{$0$} -- (8.7,-0.6)node[anchor=north]{$L$}node[anchor=west]{$z$};
        \draw[thick,->] (8.9,0.0)node[anchor=north]{$0$} -- (8.9,3.5)node[anchor=west]{$h$}node[anchor=south]{$y$};
        \draw[pattern=north east lines, pattern color=brown!50] (0,0.0) rectangle (8.7,0.5);
        \fill[white] (0,0.5) rectangle (8.7,3.5);
        \draw (0,0.5) rectangle (8.7,3.5);

        \draw[pattern=north west lines, pattern color=orange!60]  (0,2.5) -- (0,0.5) -- (8.7,0.5) -- (8.7,2.5)-- cycle;

        \draw[->]  (-0.2,1.5) -- node[left=6pt]{$N_{s,in}$} (0.2,1.5);
        \draw[->]  (8.6,1.5) -- (9.2,1.5)node[right]{$N_{s,out}$};

        
        \draw (5.25,-0.7) -- (5.25,-0.5) node[anchor=south]{$z_{k}$};
        \draw[dashed] (4.5,0) -- (4.5,3.5);
        \draw (4.5,-0.7) -- (4.5,-0.5) node[anchor=south]{$z_{k-\frac{1}{2}}$};
        \draw[dashed] (6.0,0) -- (6.0,3.5);
        \draw (6.0,-0.7) -- (6.0,-0.5)node[anchor=south]{$z_{k+\frac{1}{2}}$};
        \draw[<->] (4.5,-0.7) -- node[anchor=north]{$\Delta z$} (6.0,-0.7);

        \draw[thick,->] (5.25,3.2) -- node[anchor=south west]{$\Tilde{H}_{a,k}$} node[anchor=south east]{$N_{a,k}$} (5.25,3.8);
        \draw[thick,->] (4.05,3.0) -- node[anchor=south east]{$\Tilde{H}_{a,k-\frac{1}{2}}$} node[anchor=north east]{$N_{a,k-\frac{1}{2}}$} (4.7,3.0);
        \draw[thick,->] (5.75,3.0) -- node[anchor=south west]{$\Tilde{H}_{a,k+\frac{1}{2}}$} node[anchor=north west]{$N_{a,k+\frac{1}{2}}$} (6.2,3.0);
        \draw[thick,->] (5.25,-0.2) -- node[anchor=south west]{$\Tilde{H}_{a,in,k}$} node[anchor=south east]{$N_{a,in,k}$} (5.25,0.9) ;
        
         \draw[thick,->] (4.3,1.5)node[anchor=east]{$\Tilde{H}_{s,k-\frac{1}{2}}$} -- node[anchor=north east]{$N_{s,k-\frac{1}{2}}$}node[anchor=south east]{$\Tilde{Q}_{s,k-\frac{1}{2}}$} (4.7,1.5);
         \draw[thick,->] (5.8,1.5) -- node[anchor=north west]{$N_{s,k+\frac{1}{2}}$} node[anchor=south west]{$\Tilde{Q}_{s,k+\frac{1}{2}}$} (6.3,1.5)node[anchor=west]{$\Tilde{H}_{s,k+\frac{1}{2}}$};

        \node[anchor=south west,inner sep=0] at (0.1,0.1){Grate Belts};
        \node[anchor=south west,inner sep=0] at (0.1,1.1){Clinker};
        \node[anchor=south west,inner sep=0] at (0.1,3.1){Air};
    \end{tikzpicture}
     }
    \caption{Axial profile, with segment notation for the $k$-th volume.}
    \label{fig:axial}
\end{subfigure}
\caption{Cooler axial profile.}
    \label{fig:my_label}
\end{figure}

In clinker production, the main purpose of the cooler is to stabilize the produced cement clinker by rapid cooling (quenching). As part of the clinker production process, Belite (dicalicium silicate, \ce{C2S}) reacts above 1300$^\circ$C to form Alite (tricalicium silicate, \ce{C3S}), a main source of early cement strength. If the clinker cooling is too slow, 
Alite converts back into Belite.
\begin{subequations}
\begin{alignat}{3}\label{eq:c2stoc2s}
\ce{C3S} \rightarrow \ce{CaO} + \ce{C2S}.
\end{alignat}
\end{subequations}
A second purpose of the cooler is the recirculation of heat to improve fuel efficiency of the entire production. 
Several cooler types exist, e.g. rotary, planetary, shaft or grate belts \cite{bhatty2010innovations}.
Fig. \ref{fig:cooler} illustrates how clinker and air move in a grate belt cooler. The clinker enters from the kiln and is transported through the cooler by a grate belt. Cool air is blown through the grate belt from below to cool the clinker, the air is afterward either released or recirculated back into the process as secondary air (2nd air) and tertiary air (3rd air) flows for the heating of the kiln and calciner sections.
The entering clinker mass typically has a temperature of 1300-1450$^\circ$C, while the exiting clinker temperature is around 100-150$^\circ$C \cite{bhatty2010innovations}, above the ambient temperature
.
The Alite decomposition into Belite in \eqref{eq:c2stoc2s} happens between 900$^\circ$C and 1250$^\circ$C \cite{CS3toCS2}.
\section{A mathematical cooler model}\label{sec:CoolerModel}
For modeling the cooler as an index-1 DAE system, we consider the molar concentrations of each compound, $C$, and the internal energy densities of each phase, $\hat{U}$, as the states, $x$.
The algebraic variables, $y$, are the phase temperatures, $T$, and the pressure, $P$.
The resulting model reads:
\begin{subequations}\label{eq:dyn}
\begin{align}
    \partial_tx& = f(x,y;p),\quad x=[C;\hat U],\\
    0 &= g(x,y;p),\quad y=[T;P],
\end{align}
\end{subequations}
with $p$ being the system parameters.
The model considers two phases, the solid clinker materials and the air given by the subscripts $_s$ and $_a$. The model is formulated using a systematic modeling methodology that integrates A) algebraic relations, B) mass and energy balances,  C) thermo-physical properties, D) stoichiometry and kinetics, and E) transport phenomena.

We use a finite-volume approach to describe the cooler in nv segments of length $\Delta z = L/nv$ along its length.
Fig. \ref{fig:axial} illustrates how the cooler model is segmented, with a total height $h$, width $w$, and length $L$. We define the molar concentration vector, C, as mole per segment volume $V_\Delta(k)=w h \Delta z$, and assume all gasses are ideal.

The standard cement chemist notation is used for the compounds:
 \ce{(CaO)_2SiO_2} as \ce{C_2S}, \ce{(CaO)_3SiO_2} as \ce{C_3S}, \ce{(CaO)_3Al_2O_3} as \ce{C_3A} and \ce{(CaO)_4(Al_2O_3)(Fe_2O_3)} as \ce{C_4AF}, where C = \ce{CaO},   A = \ce{Al_2O_3}, S = \ce{SiO_2}, and  F = \ce{Fe_2O_3}. 

We utilize the following assumptions:
1) the temperatures and pressure are homogeneously within each phase;  
2) the dynamics along the width is ignored (2D); 
3) no dust entrapped in the air;
4) thermal interactions with the environment and walls are ignored; and
5) only the 6 main clinker formation reactions are included.

\subsection{Algebraic relations}
The model consists of two sets of algebraic relations. 1) Thermo-physical relations between the specific energies, $\hat U$, and temperature, pressure, and concentration of each phase:
\begin{small}
\begin{subequations}
\label{eq:EnergyAlgebra}
\begin{align}
    \hat{U}_a &= \hat{H}_a - P\hat{V}_a = H_g(T_a,P,C_a) - P 
    V_g(T_c,P,C_g), \\
    \hat{U}_s &= \hat{H}_s = H_s(T_c,P,C_s).
\end{align}
\end{subequations}
\end{small}
2) A geometric relation between the total specific volume of the gas and the solid phases: 
\begin{align}
    \hat{V}_{a} + \hat{V}_{s} = V_a(T_a,P,C_a) + V_s(T_s,P,C_s) = \hat{V}_{\Delta} = 1.
\end{align}

\subsection{Mass and Energy balances}
The compound-specific mass balances of the solid and the air phases are
\begin{subequations}\label{eq:mass} 
    \begin{align}
        \partial_tC_{s,i} &= -\nabla \vdot N_{s,i} + R_{s,i},\\
        \partial_tC_{a,i} &= -\nabla \vdot N_{a,i} + R_{a,i}, 
    \end{align}    
\end{subequations}     
where $i$ indicates the compound, $N_{j,i}$ is the molar flux of phase $j$, $R_{j,i}$ is the production rate, and the divergence $\nabla\vdot$ covers the dimensions $y$ and $z$. 

The energy balances of the solid and air phases are 
\begin{align}
    \partial_t\hat{U}_s &= -\nabla \vdot (\Tilde{H}_s + \Tilde{Q}_s)  - \hat{Q}^{rad}_{sa} - \hat{Q}^{cv}_{sa} - J_{sa}, \\
    \partial_t\hat{U}_a &= -\nabla \vdot( \Tilde{H}_a + \Tilde{Q}_a ) + \hat{Q}^{rad}_{sa}+\hat{Q}^{cv}_{sa} + J_{sa},
\end{align}
where $\Tilde{H}_{j,i}$ is the enthalpy flux, $\Tilde{Q}_j$ is the thermal conduction, $J_{sa}$ is the phase transition term and the specific heat transfer of radiation and convection is noted by $\hat{Q}^{rad}_{sa}$ and $\hat{Q}^{cv}_{sa}$.

\subsection{Thermodynamic}
Using thermo-physical models, we describe the enthalpy and volume of each phase by the functions $H(T,P,n$) and $V(T,P,n)$. For the mole vector, $n$, the models are homogeneous of order 1, and given as
\begin{small}
 \begin{align}
    H(T,P,n) &= \sum_i n_i \left(\Delta H_{f,i}(T_0,P_0) + \int^T_{T_0}c_{p,i}(\tau)d\tau \right),
    \\
    V(T,P,n) &= \begin{cases} \frac{1}{1-\eta}\sum_i n_i \left( \frac{M_i}{\rho_i} \right), \quad &\text{solid},\\ \sum_{i} n_i \left( \frac{RT}{P} \right), \quad &\text{air}.
  \end{cases}    
\end{align}
\end{small} 
$\Delta H_{f,i}(T_0,P_0)$ is the formation enthalpy at standard conditions $(T_0,P_0)$. $\eta$ is the porosity of the clinker batch, giving the bulk volume.
Assuming the clinkers are ideal spheres of equal size, the range of porosity \cite{hales_adams_bauer_dang_harrison_hoang_kaliszyk_magron_mclaughlin_nguyen_etal._2017} is
\begin{align}
    1-\frac{\pi}{3\sqrt{2}} \leq \epsilon \leq 1.
\end{align}
In the literature, a constant porosity of 0.4 is used \cite{CUI20171297}, though a more accurate porosity would depend on the clinker composition, e.g. sulfur content.
The enthalpy density, $\hat{H}$, and volume density, $\hat{V}$, can be computed as
\begin{align}
    \hat{H}_s &= H(T_s,P,C_s),\quad  &\hat{V}_s &= V(T_s,P,C_s)\\ 
    \hat{H}_a &= H(T_a,P,C_a),\quad &\hat{V}_a &= V(T_a,P,C_a). 
\end{align}

\subsection{Stoichiometry and kinetics}
The production rates $R$ in (\ref{eq:mass}) are provided by the reaction rate vector $r = r(T,P,C)$ and the stoichiometric matrix, $\nu$:
\begin{align}
    R = \begin{bmatrix}
        R_s\\R_a
    \end{bmatrix}&= \nu^Tr.
\end{align}
$R_s$ and $R_a$ are the production rate vectors of the clinker and air compounds respectively.
The following reactions are considered:
\begin{subequations}
\begin{alignat}{5}
    \text{$\#1$: }& & \ce{CaCO3} &\rightarrow \ce{CO2} + \ce{CaO}, \ & r_1,\\
    \text{$\#2$: }& & 2\ce{CaO} + \ce{SiO_2} &\rightarrow \ce{C_2S}, \ & r_2,\\
    \text{$\#3$: }& &\ce{CaO} + \ce{C_2S}&\rightarrow \ce{C_3S}, \ & r_3,\\
    \text{$\#4$: }& &3\ce{CaO} + \ce{Al_2O_3}&\rightarrow \ce{C_3A}, \ & r_4,\\
    \text{$\#5$: }& &4\ce{CaO} + \ce{Al_2O_3} + \ce{Fe_2O_3}&\rightarrow \ce{C_4AF}, \ & r_5,\\
    \text{$\#6$: }& &\ce{C_3S}&\rightarrow \ce{C_2S} + \ce{CaO}, \ & r_6,
    \end{alignat}
\end{subequations}
The rate function $r_j(T,P,C)$ is formulated as
\begin{align}
    r_j = k_r(T)\prod_lC_l^{\alpha_l},\quad k(T) = k_re^{-\frac{E_{A}}{RT}}.
\end{align}
 $k(T)$ is the Arrhenius equation. $C_l$ are the concentrations (mol/L). $\alpha_l$ are the stoichiometric values. Table \ref{tab:reaction} shows the parameters for the kinetic expressions found in the literature.
 The heat transfer for phase transition, $J_{sa}$, is given as
\begin{align}
    J_{sa} &= H(T_s,P,r_1).
\end{align}
\begin{table}
\centering
    \caption{Reaction rate coefficients.}
    \begin{tabular}{c c|c c | c c c }
     & & $k_r$ & $E_{A}$ & $\alpha_1$ & $\alpha_2$ & $\alpha_3$\\ \hline \ & & & && &\\[-1em]
    Reactions&Units & $[r_i]\cdot [C]^{-\Sigma\alpha}$ & $\frac{\text{kJ}}{\text{mol}}$& 1& 1& 1\\ \ & & & && &\\[-1em] \hline \ & & & && &\\[-1em]
    $r_1$  & $\frac{\text{kg}}{\text{m}^3s}$& $10^{8}$& $175.7$& 1& &\\\ & & & && &\\[-1em]
    $r_2$ & $\frac{\text{kg}}{\text{m}^3s}$ &$10^{7}$ & $240$& 2& 1&\\\ & & & && &\\[-1em]
    $r_3$ & $\frac{\text{kg}}{\text{m}^3s}$ & $10^{9}$& $420$& 1& 1&\\\ & & & && &\\[-1em]
    $r_4$ & $\frac{\text{kg}}{\text{m}^3s}$ &$10^{8}$ & $310$& 3& 1&\\\ & & & && &\\[-1em]
    $r_5$ & $\frac{\text{kg}}{\text{m}^3s}$ & $10^{8}$& $330$& 4& 1&1\\ & & & && &\\[-1em]
    $r_6$ & s$^{-1}$ & $0.09^*$ & $96.58^*$ & 1& & \\ \hline
    \end{tabular} 
    
    \footnotesize{The reported units and coefficients of $r_1$-$r_5$ are from \cite{Mastorakos1999CFDPF}. $^*$ estimated coefficients of $r_6$ using least-square and the Jander-data in \cite{CS3toCS2}.}
       \label{tab:reaction}
\end{table}

\subsection{Transport}
For each phase, mass is transported by advection, while energy is transported by
advection, conduction, convection, and radiation.

\subsubsection{Material flux} 
The mass transfer of each compound is driven by advection. The molar fluxes are
\begin{align}
    N_{s,i} = v_{s}C_{s,i},\quad    N_{a,i} = v_{a}C_{a,i}.
\end{align}
$v_{j}$ is the velocity vector.



\subsubsection{Enthalpy and heat transport}
The transport of energy includes the transport of enthalpy (advection) and thermal conduction (diffusion). The enthalpy flux, $\Tilde{H}$, can be computed by
\begin{align}
    \Tilde{H}_s &= H(T_s,P,N_s),\quad  &\Tilde{H}_a = H(T_a,P,N_a).
\end{align}

The thermal conduction is given by Fourier's law \cite{Perry} 
\begin{align}
    \Tilde{Q}_s = -k_s\nabla T_s,\quad & \Tilde{Q}_a = -k_a\nabla T_a
\end{align}
with $k_j$ being the thermal conductivity of the phase.

\subsubsection{Velocities}
Both the advection term of the material flux and the heat flux depend on the velocities in the cooler.

For the clinkers, we assume no dust is lifted by the airflow, so vertical movement is zero. The horizontal movement is given by the grate belt carrying the clinkers 
\begin{align}
    v_{s}(z) = [v_{s,y},v_{s,z}]^T = [0,v_{grate}(z)]^T.
\end{align}

For the air, the movement is driven by pressure differences and assumed independent of clinker and belt movements. We assume velocity is identical for each compound and is below 0.2 Mach in each direction. The velocities are given by the turbulent Darcy-Weisbach equation \cite{Darcy-Howel}:
\begin{align}
    v_{a}(y,z) &= [v_{a,y},v_{a,z}]^T,\\
    v_{a,i} &= \Big(\frac{2}{0.316}\sqrt[4]{\frac{D_{H,i}^{5}}{\mu_a\rho_a^3}}\frac{|\Delta P_i|}{\Delta i}\Big)^{\frac{4}{7}}\text{sgn}\Big(-\frac{\Delta P_i}{\Delta i}\Big),
\end{align}
derived using the Darcy friction factor, $f_D = 0.316Re^{-\frac{1}{4}}$. $\mu_a$ is the air viscosity. $\rho_a$ is the air density,
\begin{align}
    \rho_a = \frac{1}{\Hat{V}_a}\sum_iM_iC_{a,i}.
\end{align}
$D_{H,i}$ is the hydraulic diameter for a non-uniform and non-circular cross-section channel \cite{HESSELGREAVES20171},
\begin{align}
   D_{H,y} &= \frac{4V_a}{2 A_{yz}+2 A_{wy}+A_c},\ A_{wy}= w\Delta y,\\
   D_{H,z} &= \frac{4V_a}{2 A_{yz}+2 A_{wz}+A_c}, \  A_{wz}= w\Delta z,\\
   A_{yz} &= \Delta z\Delta y, \  A_c = \frac{V_s}{\frac{\pi}{6} D_p^3} (\pi D_p^2).
\end{align}
$A_c$ is the total surface area of the clinkers.

\subsubsection{Heat transfer between phases}
The convective heat transfer between phases is given by Newton's law of heat transfer,
\begin{align}
    \hat{Q}^{conv}_{sa}& = \hat{A}\beta (T_s-T_a).
\end{align}
The heat transfer due to thermal radiation is given by \cite{CUI20171297},
\begin{align}
    \hat{Q}^{rad}_{sa} &=   \hat{A}\sigma (\epsilon_sT_s^4-\epsilon_aT_a^4).
\end{align}
$\hat{A}$ is the specific surface area. $\sigma$ is the Stefan-Boltzmann's constant. $\epsilon_s$ and $\epsilon_a$ are the emissivity of each phase. $\beta$ is the heat transfer coefficient \cite{CUI20171297}:
\begin{align}
    \hat{A} &=\frac{6}{D_p}(1-\hat{V}_a), \quad    \beta = \frac{k_sNu}{D_p + 0.5\phi D_pNu}.
\end{align}
$\phi = 0.25$ is the clinker shape correction factor. $D_p$ is the average clinker particle diameter, typically around 40 mm. The Nusselt Number $Nu$ is given by
\begin{align}
    Nu &= 2+1.8Pr^{\frac{1}{3}}Re^{\frac{1}{2}},\\
    Pr &= \frac{c_{ps}\mu_a}{k_a}, \quad    Re = \frac{\rho_av_yD_p}{\mu_a}.
\end{align}
$Pr$ and $Re$ is the Prandtl and Reynold number \cite{CUI20171297}. $\mu_a$ is the air viscosity. The heat capacity $c_{ps}$ is given by:
\begin{align}   
c_{ps} &= \sum_in_ic_{m,i}.
\end{align}
$c_m$ is the molar heat capacity, shown in Table \ref{tab:heatCap}.

The emissivity of the clinker, $\epsilon_s$, can be assumed similar to the emissivity of clinker materials reported for the kiln, $\epsilon_s=0.9$ \cite{Hanein2017}.
The emissivity of the air, $\epsilon_a$, is computed using the WSGG model of 4 grey gases \cite{Johanson2011},
\begin{align}
    \epsilon_g &= \sum^4_{j=0}a_j(1-e^{-k_jS_mP(x_{H2O}+x_{CO2})}),\\
    a_0 &= 1-\sum^4_{j=1}a_j,\quad a_j=\sum^3_{i=1}c_{j,i}(\frac{T}{T_{ref}})^{i-1},\\
    k_j &= K_{1,j}+K_{2,j}\frac{x_{H2O}}{x_{CO2}}, \quad x_i = \frac{n_i}{n_g},\\
    c_{j,i} &= C_{1,j,i}+C_{2,j,i}\frac{x_{H2O}}{x_{CO2}}+C_{3,j,i}(\frac{x_{H2O}}{x_{CO2}})^2.
\end{align}
$T_{ref}$ is 1200K. $x_i$ is the molar fraction of \ce{CO2} or \ce{H2O}.
The $K$ and $C$ coefficients are given by Table 1 in \cite{Johanson2011}.
$S_m$ is the total path length.
In this paper, $S_m$ is taken as the distance between the centroids of the solid and gas phases, $h/2$, as an average length.


\subsubsection{Viscosity and conductivity} 
In \cite{Sutherland1893}, the correlation between temperature and viscosity of a pure gas is given by
\begin{align}
    \mu_{a,i}(T) = \mu_0 \left(\frac{T}{T_0}\right)^{\frac{3}{2}}\frac{T_0+S_{\mu,i}}{T+S_{\mu,i}}.
\end{align}
Table \ref{tab:Data-Coeff-gas} provides the two measures of viscosity for calibrating $S_{\mu,i}$. 
A correlation between the air concentrations and the viscosity and thermal conductivity of a gas mixture is provided by Wilke's formula \cite{Wilke1950} and the Mason-Saxena modified Wassiljewa's equation \cite{Poling2001Book},
\begin{subequations}
\begin{align}
    \mu_a &= \sum_i\frac{x_i\mu_{a,i}}{\sum_jx_j\phi_{ij}},\quad k_a = \sum_i\frac{x_ik_{a,i}}{\sum_jx_j\phi_{ij}},\\
    \phi_{ij} &= \bigg(1+\sqrt{\frac{\mu_{a,i}}{\mu_{a,j}}}\sqrt[4]{\frac{M_j}{M_i}}\bigg)^2\bigg(2\sqrt{2}\sqrt{1+\frac{M_i}{M_j}}\bigg)^{-1}.
\end{align}
\end{subequations}
$x_i$ being the mole fraction of component $i$

The thermal conductivity of the solid phase depends on the clinker composition and the porosity. Assuming the compounds can be considered as layers, the thermal conductivity is then given by the serial thermal conductivity \cite{Perry},
\begin{align}
    \frac{1}{k_s} = \eta\frac{1}{k_{a}} + (1-\eta)\sum_i\frac{V_{s,i}}{V_s}\frac{1}{k_{s,i}},
\end{align}
where the width ratio is represented by a volume ratio. 

Table \ref{tab:Data-Coeff-solid} and \ref{tab:Data-Coeff-gas} show the material data of each compound.
\subsection{boundary conditions}
The model is bounded by the following variables. 1) the molar flux of air blown in by the fans below the grate belt; 2) the molar influx of clinker compounds; 3) the external pressures; the kiln, 3rd air duct, and environment pressure; 4) the location of the connections to neighboring units; i.e. which segment is connected to which external pressure.
\begin{figure}
    \centering
    \includegraphics[width=\linewidth, trim={4.5cm, 5.5cm, 8.5cm, 6cm},clip]{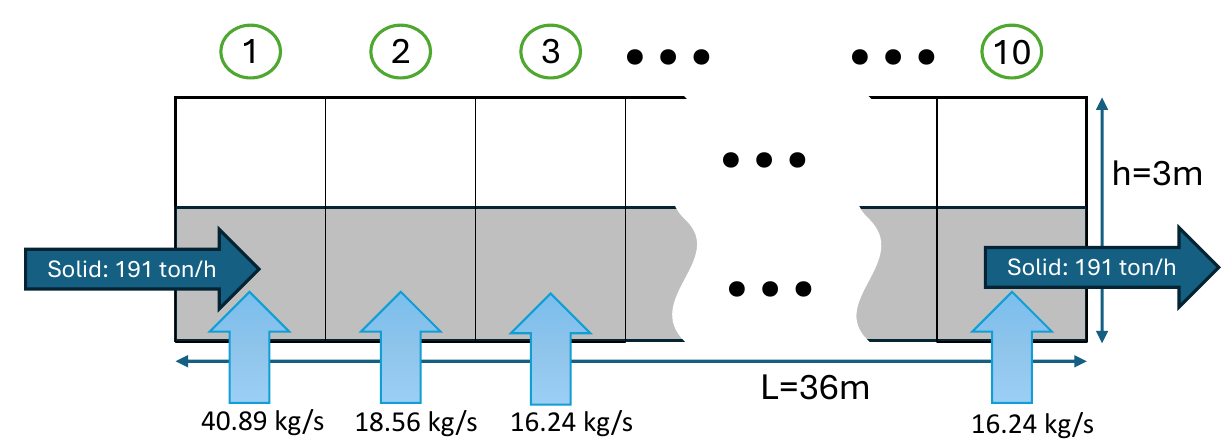}
    \caption{Segment layout of the cooler with a selection of flows; solid flows (dark blue) and gas flows (light blue).}
    \label{fig:diagram}
\end{figure}

\begin{figure*}[ht]
    \centering
    \begin{subfigure}[b]{\textwidth}
    \includegraphics[width=\textwidth,trim={0.9cm, 8.7cm, 1.1cm, 8.2cm},clip]{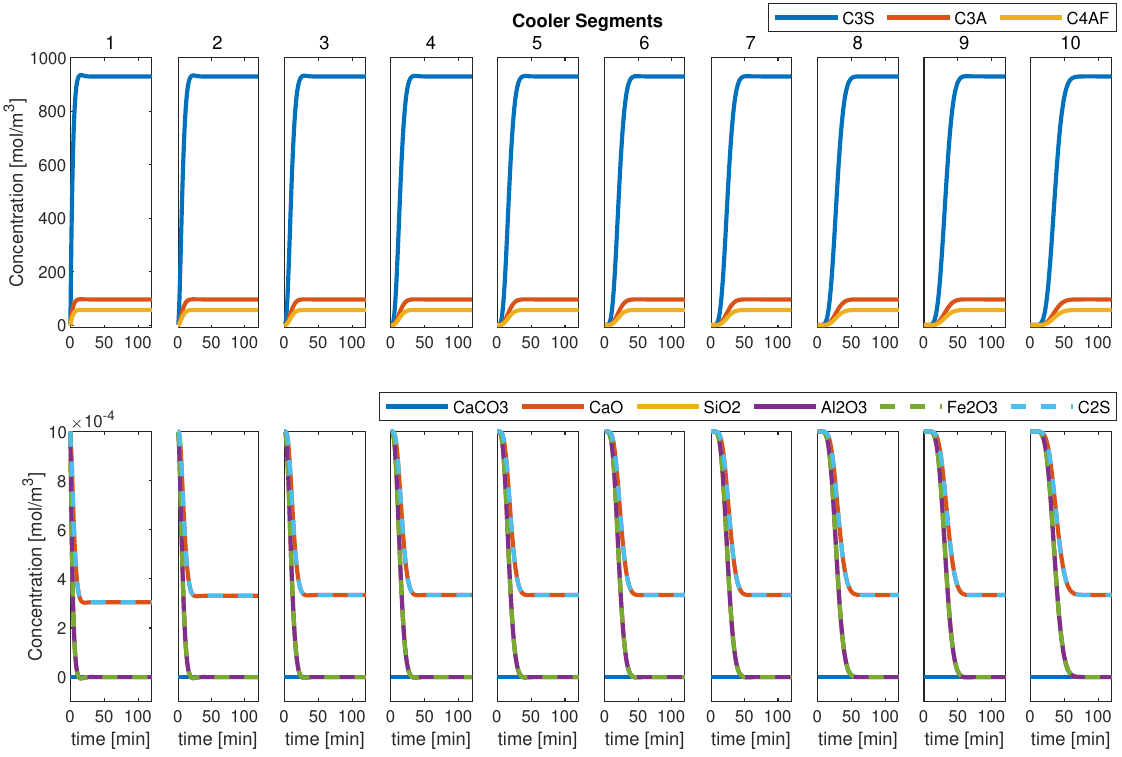}
    \caption{ Evolution of the concentration of clinker components in the cooler.}
    \label{fig:DynS1}        
    \end{subfigure}
    
    \begin{subfigure}[b]{\textwidth}
    \centering
     \includegraphics[width=\textwidth,trim={0.9cm, 11.7cm, 1.0cm, 11.7cm},clip]{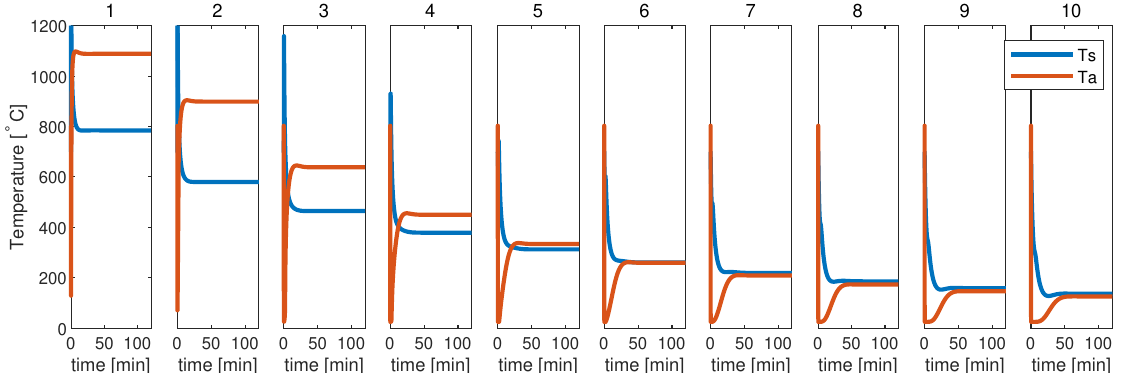}
    \caption{ Evolution of the solid and air temperatures in the cooler.}
    \label{fig:DynT}
    \end{subfigure}
     \caption{ Dynamic simulation of cooler concentration and temperature. The cooler is divided into 10 segments of finite volumes. 
     }\label{fig:Dyn}
\end{figure*}

\section{Simulation Results}\label{sec:SimulationResults}
To demonstrate the simulation model, we simulate the cooler for 2 hours of operation.
A steady-state simulation from FLSmidth Cement is used for comparison.
We consider a 36 m long cooler with 4 m width and 3 m tall. We use a 10-segment model, as illustrated in Fig. \ref{fig:diagram}.
The operation is set to a 191 ton/h clinker production. 
The clinker inflow composition is 
79.96\% \ce{C3S}, 9.76\% \ce{C3A}, and 10.28 \% \ce{C4AF} with a temperature of 1450$^\circ$C. 
The grate belts were run uniformly with a residence time of 36 min, 0.017 m/s.
The air inflows for each segment are 40.89 kg/s for the first, 18.56 kg/s for the second, and 16.24 kg/s for the remaining segments. The inflow is regular air with 1\% \ce{H2O} vapor at 25$^\circ$C.
The ambient pressure is 1.01325 bar (1 atm), the kiln pressure is 1.0115 bar, and the 3rd air pressure is 1.01125 bar. The kiln pressure is set on segment 1, the 3rd air pressure is set on segment 2, and the ambient pressure covers the remaining segments.
We used a uniform clinker concentration of 10 $\text{mol}/\text{m}^3$ at 700$^\circ$C for initialization, with the air being 800$^\circ$C.
The specific scenario was chosen to clearly highlight the dynamic aspects of the model, e.g. \ce{C3S} decomposition, thus the 79.96\% \ce{C3S} complete reaction scenario instead of the typical 60-70\% range with the corresponding \ce{C2S} and \ce{CaO} present.


\subsection{Model calibration}
The Darcy friction factor $f_D$ is manually calibrated to fit the expected pressure drop in the reference simulation.
To obtain a drop of 1.06 bar to 1.015 bar, a factor of 100 is used; representing a less smooth scenario of gas movement.

\subsection{Dynamic performance}
Fig. \ref{fig:Dyn} illustrates the dynamic behavior of the model.
Fig. \ref{fig:DynS1} shows the solid molar concentrations and Fig. \ref{fig:DynT} shows the temperature. 
We observe the system settles to a steady state after 50-60 min, with the first segment settling around 20 min. 
The concentrations increase monotonically as the clinker flows in except \ce{C3S}, which has slight overshoots of up to 2 mol/m$^3$ due to the decomposition of Alite. 
We observe that the initial temperatures in Fig. \ref{fig:DynT} drop rapidly and then increase as the material influx wave arrives.

\begin{figure}
    \centering
     \includegraphics[width=0.5\textwidth,trim={3.4cm, 9.5cm, 3.5cm, 9.5cm},clip]{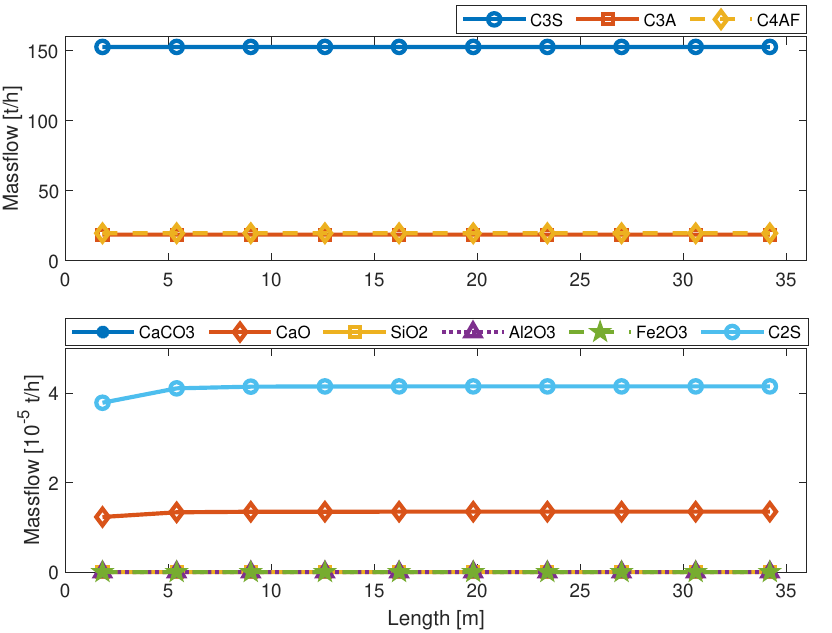}
    \caption{ The steady-state mass flows of solids along the length of the cooler. All of the decomposition of Alite (\ce{C3S}) to Belite (\ce{C2S}) is quickly halted by the cooling.  }
    \label{fig:Smassflow}
\end{figure}

\begin{figure}
    \centering
    \includegraphics[width=0.5\textwidth,trim={3.2cm, 9.4cm, 4cm, 10cm},clip]{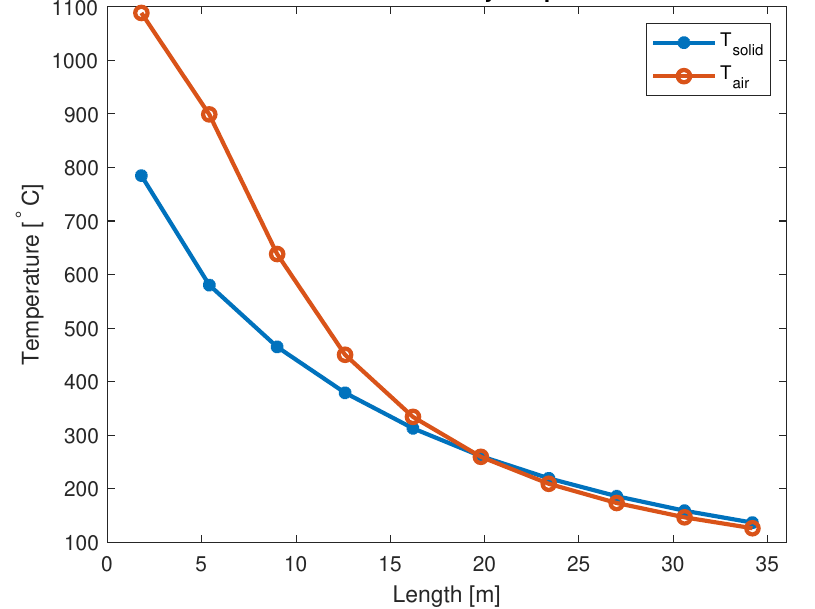}
    \caption{The steady-state temperature of the clinker mass and air content in the cooler.}
    \label{fig:Temp}
\end{figure}

\begin{figure}
    \centering
    \includegraphics[width=0.5\textwidth,trim={3.5cm, 11.3cm, 4cm, 11.5cm},clip]{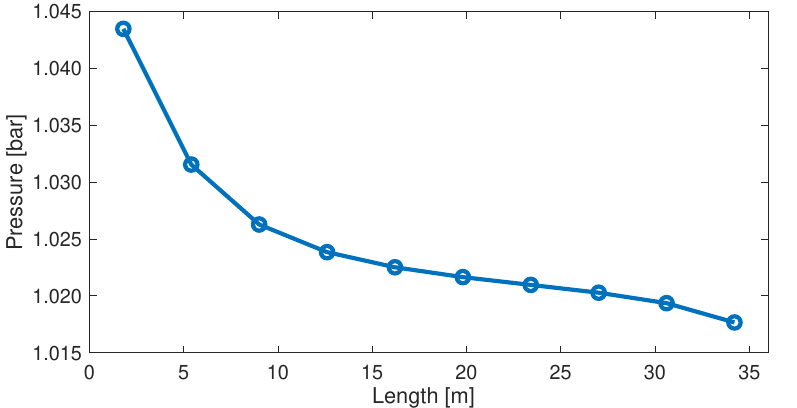}
    \caption{The steady-state pressure profile in the cooler.}
    \label{fig:press}
\end{figure}
\subsection{Steady-state results}
Fig. \ref{fig:Smassflow} shows the steady-state mass flow of the clinker compounds.
The mass flow of \ce{C2S} and \ce{CaO} shows slight increases from their zero inflow, indicating \ce{C3S} decomposition. 
Confirming the \ce{C3S} overshoot is due to decomposition.
Fig. \ref{fig:Temp} shows the steady-state temperature profile of the cooler.
The air temperature of segment 1 is 1088.7$^\circ$C matching the 1089$^\circ$C for 2nd air flows. The air temperature of segment 2 is 899.1$^\circ$C comparable to the 968$^\circ$C for 3rd air flows.
The solid temperature at segment 1 is 784.8$^\circ$C below the decomposition range. In segment 2, the temperature at 580.3$^\circ$C is well below the range, thus halting the \ce{C3S} decomposition.
By extrapolating the slope of the solid temperature, the outlet clinker temperature is estimated to 125.3$^\circ$C, which fits within the temperature range of the outlet clinker, 100 - 150$^\circ$C \cite{bhatty2010innovations}.
Fig. \ref{fig:press} shows the steady-state pressure profile.
The pressure of segment 1 lies in between the kiln pressure and inflow pressure (1.06 bar), with the pressure throughout the cooler decreasing towards the ambient pressure.


\section{Conclusion}\label{sec:Conclusion}
In this paper, we modeled a cooler in a cement plant's pyro-section as an index-1 DAE model.
The model is constructed using a systematic modeling approach involving thermo-physical properties, transport phenomena, stoichiometry and kinetics, mass and energy balances, and algebraic relations for the volume and internal energy.
A model simulation illustrating the cooler dynamics is provided, including the cooling aspects and the decomposition of Alite.
The simulation is shown to match the reference operation qualitatively.  

The utility of the cooler model covers both dynamic and steady-state simulation.
For dynamic simulations of the full pyro-section, the cooler model will be connected to modules for the rotary kiln, the calciner, and the pre-heating cyclones.
The model allows for the prediction and evaluation of operation strategies and serves as a basis for improving clinker quality and energy efficiency through optimization, i.e., the cooling rate.

\bibliographystyle{IEEEtran}
\bibliography{IEEEabrv,biblio}

\begin{thebibliography}{10}
\providecommand{\url}[1]{#1}
\csname url@rmstyle\endcsname
\providecommand{\newblock}{\relax}
\providecommand{\bibinfo}[2]{#2}
\providecommand\BIBentrySTDinterwordspacing{\spaceskip=0pt\relax}
\providecommand\BIBentryALTinterwordstretchfactor{4}
\providecommand\BIBentryALTinterwordspacing{\spaceskip=\fontdimen2\font plus
\BIBentryALTinterwordstretchfactor\fontdimen3\font minus
  \fontdimen4\font\relax}
\providecommand\BIBforeignlanguage[2]{{%
\expandafter\ifx\csname l@#1\endcsname\relax
\typeout{** WARNING: IEEEtran.bst: No hyphenation pattern has been}%
\typeout{** loaded for the language `#1'. Using the pattern for}%
\typeout{** the default language instead.}%
\else
\language=\csname l@#1\endcsname
\fi
#2}}

\bibitem{CO2Techreport}
J.~Lehne and F.~Preston, ``Making concrete change: Innovation in low-carbon
  cement and concrete,'' Chatham House, Tech. Rep., june 2018.

\bibitem{METZGER1983491}
M.~Metzger, ``Simplified mathematical model of the rotary kiln unit dynamical
  properties,'' \emph{IFAC Proceedings Volumes}, vol.~16, no.~10, pp. 491--497,
  1983.

\bibitem{CUI20171297}
Z.~Cui, W.~Shao, Z.~Chen, and L.~Cheng, ``Mathematical model and numerical
  solutions for the coupled gas–solid heat transfer process in moving packed
  beds,'' \emph{Applied Energy}, vol. 206, pp. 1297--1308, 2017.

\bibitem{MUJUMDAR2007}
K.~S. Mujumdar, K.~Ganesh, S.~B. Kulkarni, and V.~V. Ranade, ``Rotary cement
  kiln simulator ({RoCKS}): Integrated modeling of pre-heater, calciner, kiln
  and clinker cooler,'' \emph{Chem. Eng. Sci.}, vol.~62, no.~9, pp. 2590--2607,
  2007.

\bibitem{Svensen2024Kiln}
J.~L. Svensen, W.~R.~L. da~Silva, J.~P. Merino, D.~Sampath, and J.~B.
  J{\o}rgensen, ``A dynamical simulation model of a cement clinker rotary
  kiln,'' in \emph{European Control Conference 2024}, 2024, pp. 1--7.

\bibitem{Svensen2024Calciner}
J.~L. Svensen, W.~R.~L. da~Silva, and J.~B. J{\o}rgensen, ``A first-engineering
  principles model for dynamical simulation of a calciner in cement
  production,'' in \emph{12th IFAC Symposium on Advanced Control of Chemical
  Processes (ADCHEM 2024)}, 2024, pp. 1--7.

\bibitem{bhatty2010innovations}
J.~Bhatty, F.~Macgregor~Miller, and S.~H. Kosmatka, \emph{Innovations in
  Portland Cement Manufacturing}.\hskip 1em plus 0.5em minus 0.4em\relax
  Portland Cement Association, 2010.

\bibitem{chatterjee2018cement}
A.~Chatterjee, \emph{Cement Production Technology: Principles and
  Practice}.\hskip 1em plus 0.5em minus 0.4em\relax CRC Press, 2018.

\bibitem{bye1999portland}
G.~Bye, \emph{Portland Cement: Composition, Production and Properties}.\hskip
  1em plus 0.5em minus 0.4em\relax Thomas Telford, 1999.

\bibitem{CS3toCS2}
X.~Li, X.~Shen, M.~Tang, and X.~Li, ``Stability of tricalcium silicate and
  other primary phases in portland cement clinker,'' \emph{Ind. Eng. Chem.
  Res.}, vol.~53, no.~5, pp. 1954--1964, 2014.

\bibitem{hales_adams_bauer_dang_harrison_hoang_kaliszyk_magron_mclaughlin_nguyen_etal._2017}
T.~Hales, M.~Adams, G.~Bauer, T.~D. Dang, J.~Harrison, L.~T. Hoang,
  C.~Kaliszyk, V.~Magron, S.~Mclaughlin, T.~T. Nguyen, Q.~T. Nguyen, T.~Nipkow,
  S.~Obua, J.~Pleso, J.~Rute, A.~Solovyev, T.~H.~A. Ta, N.~T. Tran, T.~D.
  Trieu, J.~Urban, K.~Vu, and R.~Zumkeller, ``A formal proof of the kepler
  conjecture,'' \emph{Forum of Math. Pi}, vol.~5, p.~e2, 2017.

\bibitem{Mastorakos1999CFDPF}
E.~Mastorakos, A.~Massias, C.~D. Tsakiroglou, D.~A. Goussis, V.~N. Burganos,
  and A.~C. Payatakes, ``{CFD} predictions for cement kilns including flame
  modelling, heat transfer and clinker chemistry,'' \emph{Applied Mathematical
  Modelling}, vol.~23, pp. 55--76, 1999.

\bibitem{Perry}
D.~W. Green and R.~H. Perry, Eds., \emph{Perry's Chemical Engineers' Handbook},
  8th~ed.\hskip 1em plus 0.5em minus 0.4em\relax McGraw Hill, 2008.

\bibitem{Darcy-Howel}
G.~W. Howell and T.~M. Weathers, \emph{Aerospace Fluid Component Designers'
  Handbook. Vol. I, Rev. D}.\hskip 1em plus 0.5em minus 0.4em\relax TRW Systems
  Group, 1970.

\bibitem{HESSELGREAVES20171}
J.~E. Hesselgreaves, R.~Law, and D.~A. Reay, ``Chapter 1 - introduction,'' in
  \emph{Compact Heat Exchangers}, 2nd~ed.\hskip 1em plus 0.5em minus
  0.4em\relax Butterworth-Heinemann, 2017, pp. 1--33.

\bibitem{Hanein2017}
T.~Hanein, F.~P. Glasser, and M.~N. Bannerman, ``One-dimensional steady-state
  thermal model for rotary kilns used in the manufacture of cement,''
  \emph{Adv. Appl. Ceram.}, vol. 116, no.~4, pp. 207--215, 2017.

\bibitem{Johanson2011}
R.~Johansson, B.~Leckner, K.~Andersson, and F.~Johnsson, ``Account for
  variations in the \ce{H2O} to \ce{CO2} molar ratio when modelling gaseous
  radiative heat transfer with the weighted-sum-of-grey-gases model,''
  \emph{Combustion and Flame}, vol. 158, pp. 893--901, 2011.

\bibitem{Sutherland1893}
W.~Sutherland, ``The {V}iscosity of {G}ases and {M}olecular {F}orce,''
  \emph{Philos Mag series 5}, vol.~36, no. 223, pp. 507--531, 1893.

\bibitem{Wilke1950}
C.~R. Wilke, ``A viscosity equation for gas mixtures,'' \emph{The Journal of
  Chemical Physics}, vol.~18, no.~4, pp. 517--519, 1950.

\bibitem{Poling2001Book}
B.~E. Poling, J.~M. Prausnitz, and J.~P. O'Connel, \emph{The Properties of
  Gases and Liquids}.\hskip 1em plus 0.5em minus 0.4em\relax McGraw-Hill, 2001.

\bibitem{CRC2022}
J.~Rumble, Ed., \emph{CRC handbook of chemistry and physics}, 103rd~ed.\hskip
  1em plus 0.5em minus 0.4em\relax CRC Press, 2022.

\bibitem{Ichim2018}
A.~Ichim, C.~Teodoriu, and G.~Falcone, ``Estimation of cement thermal
  properties through the three-phase model with application to geothermal
  wells,'' \emph{Energies}, vol.~11, no.~10, 2018.

\bibitem{PhysRevApplied}
M.~J. Abdolhosseini~Qomi, F.-J. Ulm, and R.~J.-M. Pellenq, ``Physical origins
  of thermal properties of cement paste,'' \emph{Phys. Rev. Appl.}, vol.~3, p.
  064010, Jun 2015.

\bibitem{Du2021}
Y.~Du and Y.~Ge, ``Multiphase model for predicting the thermal conductivity of
  cement paste and its applications,'' \emph{Materials}, vol.~14, no.~16, 2021.

\bibitem{Portland}
G.~C. Bye, Ed., \emph{Portland Cement: Composition, Production and Properties},
  2nd~ed.\hskip 1em plus 0.5em minus 0.4em\relax Thomas Telford, 1999.

\bibitem{HANEIN2020106043}
T.~Hanein, F.~P. Glasser, and M.~N. Bannerman, ``Thermodynamic data for cement
  clinkering,'' \emph{Cem. Concr. Res.}, vol. 132, p. 106043, 2020.

\end{thebibliography}


\section*{APPENDIX}

\section{Material Properties}\label{app:PhysicalProperties}
Table \ref{tab:Data-Coeff-solid} and Table \ref{tab:Data-Coeff-gas} shows literature data for the material properties of the compounds for the solid and air phases. 
Table \ref{tab:heatCap} reports the parameters ($C_0$, $C_1$,$C_2$) for computing the specific heat capacities of the components \cite{Svensen2024Kiln},
\begin{equation}
c_p = C_0 + C_1 T + C_2 T^2.
\end{equation}

\begin{table}
    \centering
    \caption{Material properties of the solid phase}
    \begin{tabular}{c|ccc}
         &\shortstack{Thermal\\ Conductivity} & Density & \shortstack{Molar \\mass}\\ \hline
         & & & \\[-1em]
         Units    & $\frac{\text{W}}{\text{K m}}$ & $\frac{\text{g}}{\text{cm}^3}$ & $\frac{\text{g}}{\text{mol}}$ \\ & & &\\[-1em]\hline
         \ce{CaCO_3} &  2.248$^a$& 2.71$^b$  &100.09$^b$\\ 
         \ce{CaO}     & 30.1$^c$ &  3.34$^b$ &56.08$^b$\\ 
         \ce{SiO_2}   &  1.4$^{a,c}$& 2.65$^b$  &60.09$^b$\\ 
         \ce{Al_2O_3} &  12-38.5$^c$ 36$^a$& 3.99$^b$  &101.96$^b$\\ 
         \ce{Fe_2O_3} &  0.3-0.37$^c$& 5.25$^b$  &159.69$^b$\\ \hline
         \ce{C2S}     &  3.45$\pm$0.2$^d$& 3.31$^d$  &$172.24^g$\\ 
         \ce{C3S}     & 3.35$\pm$0.3$^d$ & 3.13$^d$ & 228.32$^b$\\ 
         \ce{C3A}     &  3.74$\pm$0.2$^e$& 3.04$^b$ & 270.19$^b$\\ 
         \ce{C4AF}    &  3.17$\pm$0.2$^e$& 3.7-3.9$^f$  &$485.97^g$ \\ \hline        
    \end{tabular}   
    
    \footnotesize{$^a$ from \cite{Perry}, $^b$ from \cite{CRC2022}, $^c$ from \cite{Ichim2018}, $^d$ from \cite{PhysRevApplied}, $^e$ from \cite{Du2021},\\ $^f$ from \cite{Portland}, $^g$ Computed from the above results} 
    \label{tab:Data-Coeff-solid}
\end{table}
\begin{table}
    \centering
    \caption{Material properties of the gas phase}
    \def\arraystretch{1.5}
    \begin{tabular}{c|ccc}
           &\shortstack{Thermal\\ Conductivity$^a$} & \shortstack{Molar\\ mass$^a$} & Viscosity$^a$\\ \hline
        Units   & $\frac{10^{-3}\text{W}}{\text{K m}}$ & $\frac{\text{g}}{\text{mol}}$ & $\mu$Pa s \\ \hline
         \ce{CO_2}    &\shortstack{\strut 16.77 (T=300K)\\ 70.78 (T=1000K) }  & 44.01  & \shortstack{\strut 15.0 (T=300K)\\ 41.18 (T=1000K) }\\\hline
         \ce{N_2} & \shortstack{\strut 25.97(T=300K)\\  65.36(T=1000K) }  &28.014 &  \shortstack{ \strut 17.89(T=300K)\\  41.54(T=1000K) } \\\hline
         
         \ce{O_2}  & \shortstack{\strut 26.49(T=300K)\\  71.55(T=1000K) } &  31.998 &  \shortstack{\strut 20.65 (T=300K)\\ 49.12 (T=1000K) }\\\hline
         
         \ce{Ar}  & \shortstack{\strut 17.84 (T=300K)\\ 43.58 (T=1000K) }& 39.948&  \shortstack{\strut  22.74(T=300K)\\  55.69(T=1000K) }\\\hline
         
         \ce{CO}  & \shortstack{\strut 25(T=300K)\\  43.2(T=600K) } & 28.010& \shortstack{\strut 17.8(T=300K)\\  29.1(T=1000K) } \\\hline
         
         
         \ce{H_2O} & \shortstack{\strut 609.50(T=300K)\\  95.877(T=1000K) }&18.015 &  \shortstack{\strut 853.74(T=300K)\\  37.615(T=1000K) }\\\hline
         
         \ce{H_2} & \shortstack{\strut 193.1 (T=300K)\\ 459.7 (T=1000K) }&2.016 &  \shortstack{ \strut 8.938(T=300K)\\ 20.73 (T=1000K) }\\\hline
    \end{tabular}

    $^a$ from \cite{CRC2022}, $^b$ from \cite{Poling2001Book}
    \label{tab:Data-Coeff-gas}
\end{table}

\begin{table}
    \centering
    \caption{Molar heat capacity}
    \begin{tabular}{c|c c  c| c}
            & $C_0$ & $C_1$ & $C_2$ & \shortstack{Temperature\\ range}\\ \hline & & & &\\[-1em]
         Units & $\frac{\text{J}}{\text{mol}\cdot\text{K}}$& $\frac{10^{-3}\text{J}}{\text{mol}\cdot \text{K}^2}$&$ \frac{10^{-5}\text{J}}{\text{mol}\cdot \text{K}^3}$ & K\\ & & & &\\[-1em] \hline
         \ce{CaCO_3}$^a$ &  23.12 &  263.4 & -19.86 & 300-600\\ 
         \ce{CaO}$^b$     &  71.69& -3.08  & 0.22  & 200 - 1800\\ 
         \ce{SiO_2}$^b$   & 58.91 &  5.02 & 0& 844 - 1800\\ 
         \ce{Al_2O_3}$^b$ &  233.004&-19.5913   &0.94441  & 200 - 1800\\ 
         \ce{Fe_2O_3}$^a$ & 103.9  & 0 & 0 & -\\ \hline
         \ce{C2S}$^b$     &  199.6& 0  &0  & 1650 - 1800\\ 
         \ce{C3S}$^b$     &  333.92&  -2.33&  0& 200 - 1800\\ 
         \ce{C3A}$^b$    & 260.58  & 9.58/2 & 0  &298 - 1800\\ 
         \ce{C4AF}$^b$  &  374.43& 36.4 & 0 &  298 - 1863\\ \hline
         \ce{CO_2}$^a$    & 25.98 &43.61 &-1.494 & 298 - 1500\\
         \ce{N_2}$^a$ & 27.31&5.19 &-1.553e-04 & 298 - 1500\\ 
         \ce{O_2}$^a$ & 25.82&12.63 &-0.3573  & 298 - 1100\\ 
         \ce{Ar}$^a$ & 20.79 & 0 & 0  & 298 - 1500\\ 
         \ce{CO}$^a$ & 26.87& 6.939  & -0.08237  & 298 - 1500\\ 
         \ce{H_2O}$^a$&30.89 & 7.858  &0.2494 & 298 - 1300\\ 
         \ce{H_2}$^a$&   28.95& -0.5839&  0.1888 & 298 - 1500\\ \hline         
    \end{tabular}
    
    \footnotesize{ $^a$ based on data from  \cite{CRC2022}, $^b$ coefficients from \cite{HANEIN2020106043} }
    \label{tab:heatCap}
\end{table}

\end{document}